\documentclass[final,3p,times,12pt]{elsarticle}
\makeatletter
\def\ps@pprintTitle{%
 \let\@oddhead\@empty
 \let\@evenhead\@empty
 \def\@oddfoot{\footnotesize\itshape
       \hfill\today}%
 \let\@evenfoot\@oddfoot}
\makeatother

\usepackage[utf8]{inputenc}
\DeclareUnicodeCharacter{E5F8}{<replacement>}
\usepackage[T1]{fontenc}
\usepackage{amssymb}
\usepackage[symbol]{footmisc}
\usepackage{xurl}
\usepackage{caption}
\usepackage{subcaption}
\usepackage{multirow}
\usepackage{xcolor}
\usepackage{amsmath}
\usepackage{xcolor}
\usepackage[section]{placeins}
\definecolor{darkgreen}{rgb}{0.0, 0.5, 0.0}

\begin{document}

\begin{frontmatter}



\title{Tuning the surface energy of fluorinated diamond-like carbon coatings via plasma immersion ion implantation plasma-enhanced chemical vapor deposition with 1,1,1,2-tetrafluoroethane}



\author[inst1]{Yuhan Tong\textsuperscript{\textsection}}
\author[inst1]{Maryam Zahedian\textsuperscript{\textsection}}
\author[inst2]{Aiping Zeng}
\author[inst1]{Ricardo Vidrio}
\author[inst2,inst3]{Mike Efremov}
\author[inst1]{Shenwei Yin}
\author[inst1]{Hongyan Mei}
\author[inst2]{Patrick Heaney}
\author[inst1]{Jennifer T. Choy}

\affiliation[inst1]{organization={Department of Electrical and Computer Engineering, University of Wisconsin Madison},
            addressline={1415 Engineering Dr}, 
            city={Madison},
            postcode={53706}, 
            state={WI},
            country={USA}}

\affiliation[inst2]{organization={NCD Technologies},
            addressline={510 Charmany Dr}, 
            city={Madison},
            postcode={53719}, 
            state={WI},
            country={USA}}

\affiliation[inst3]{organization={Wisconsin Center for Nanoscale Technology, University of Wisconsin - Madison},
            addressline={1509 University Avenue}, 
            city={Madison},
            postcode={53706}, 
            state={WI},
            country={USA}}

\begin{abstract}
\begingroup\renewcommand\thefootnote{\textsection}
\footnotetext{Equal contribution}
\endgroup

We demonstrate an environmentally friendly and scalable method to create fluorine-doped diamond-like carbon (F-DLC) coatings using plasma immersion ion implantation plasma-enhanced chemical vapor deposition (PIII-PECVD) with 1,1,1,2-tetrafluoroethane. F-DLC films tend to have low wettability and good mechanical flexibility, which make them suitable for applications in biomedical devices and antibiofouling surfaces. We report on the effects of fluorine incorporation on the surface chemistry, surface energy, and morphology of these coatings, showing that our method is effective in increasing the fluorine content in the F-DLC up to 40\%. We show that the addition of fluorine leads to a decrease in surface energy, which is consistent with a reduction in surface wettability. 
\end{abstract}


\end{frontmatter}


\section{Introduction}
\label{Introduction}

Fluorine-doped diamond-like coatings (F-DLC) provide an attractive alternative to standard DLC coatings in applications involving surfaces with lowered surface energy and increased hydrophobicity \cite{su2010modification, ishihara2006antibacterial, bendavid2010properties}. These applications include surface passivation and modification of microelectromechanical systems (MEMS) devices, integrated circuits, and implantable medical devices \cite{bendavid2010properties, hasebe2006fluorinated, zhang2015recent}, which especially benefit from the reduced internal stress and enhanced mechanical flexibility of F-DLC \cite{bendavid2010properties}.

F-DLC films have been previously deposited using various physical deposition techniques, such as radio frequency plasma-enhanced chemical vapor deposition (RF-PECVD), using a mixture of acetylene (C\(_2\)H\(_2\)) and argon (Ar) with tetrafluoromethane (CF\(_4\)) \cite{ahmed2012evaluation}, as well as pulsed cathodic arc deposition with Ar and octafluorocyclobutane (C\(_4\)F\(_8\)) gases \cite{zhang2020effect}. Moreover, plasma immersion ion implantation and deposition (PIII–D) and RF magnetron sputtering with CF\(_4\) and methane (CH\(_4\)) \cite{yao2004structural} are also employed to produce F-DLC coatings.

Different F-DLC coating techniques can considerably impact properties such as total surface free energy and surface roughness of the resulting films. For example, in a study using RF magnetron sputtering deposition, a reduction in total surface free energy from 43.8 $\mathrm{mJ/m}^2$ to 34.5 $\mathrm{mJ/m}^2$ was observed as the fluorine percentage (F\%) increased from 0 to 25\% \cite{ishihara2006antibacterial}. In contrast, in another study using plasma-activated chemical vapor deposition, the reduction in total surface free energy of the F-DLC coatings was reported to be less than 3 $\text{mJ/m}^2$ as fluorine percentage increased from 0\% to 23.1\%. In addition, surface roughness of these F-DLC coatings has been reported to either increase \cite{su2010modification, zhang2020effect, ahmed2010influence}, decrease (due to the etching effect of fluorine gas) \cite{yao2004structural} or even show the combination of both changes \cite{jiang2013effect}, depending on the specific deposition techniques, gases, and substrates used. 

Here we introduce a novel and environmentally safe method of F-DLC deposition, using PIII-PECVD with CF$_3$CH$_2$F (1,1,1,2-tetrafluoroethane) as fluorine source. PIII-PECVD is particularly notable for its ability to easily scale up deposition over larger surface areas and deposit uniform coatings on complex surfaces \cite{chu2004recent}, a feature that is especially useful for surface modifications of bio-materials and MEMS structures. Although there exists various process gases for the fluorine source, we choose to use Freon 134a or 1,1,1,2-tetrafluoroethane (CF$_3$CH$_2$F) with CH\(_4\) due to its advantages such as lower cost and wide availability as well as its more environmentally friendly nature, with a global warming potential (GWP) of 1300 \cite{Tetrafluoromethane}, which is notably lower than the GWPs of CF\(_4\) (5700) and C\(_4\)F\(_8\) (10000) \cite{suzuki2008high}. Additionally, it poses fewer risks in handling and storage compared to CF\(_4\) and C\(_4\)F\(_8\), making it a safer choice \cite{Tetrafluoromethane, suzuki2008high}. This mixture of gases has been previously used for a RF-PECVD technique \cite{nery2017produccao} but the effects of fluorine incorporation on surface energy, as well as the use of these gases in PIII-PECVD have not been reported.

We explore different F:C atomic ratios in process gas mixtures to fabricate F-DLC films with varying fluorine content. For each gas ratio, we characterize the surface wettability, surface free energy, chemical groups, surface roughness, and mechanical properties of these films. We find that fluorine incorporation significantly impacts the chemical bonding and roughness of the films, both of which affect wettability. Increasing fluorine content leads to the replacement of sp$^3$ carbon bonds by C-CF, C-F, and C-F$_2$ bonds, which  reduce the dispersive component of the surface energy and decrease wettability. Additionally, fluorine-induced etching modifies the surface roughness which provides a confounding factor that influences wettability especially for polar liquids.


\section{PIII-PECVD growth of F-DLC with 1,1,1,2-tetrafluoroethane}
\label{growth}

\begin{figure}[hbt!]
    \centering
    \begin{subfigure}{\textwidth}
        \centering
        \includegraphics[width=\textwidth]{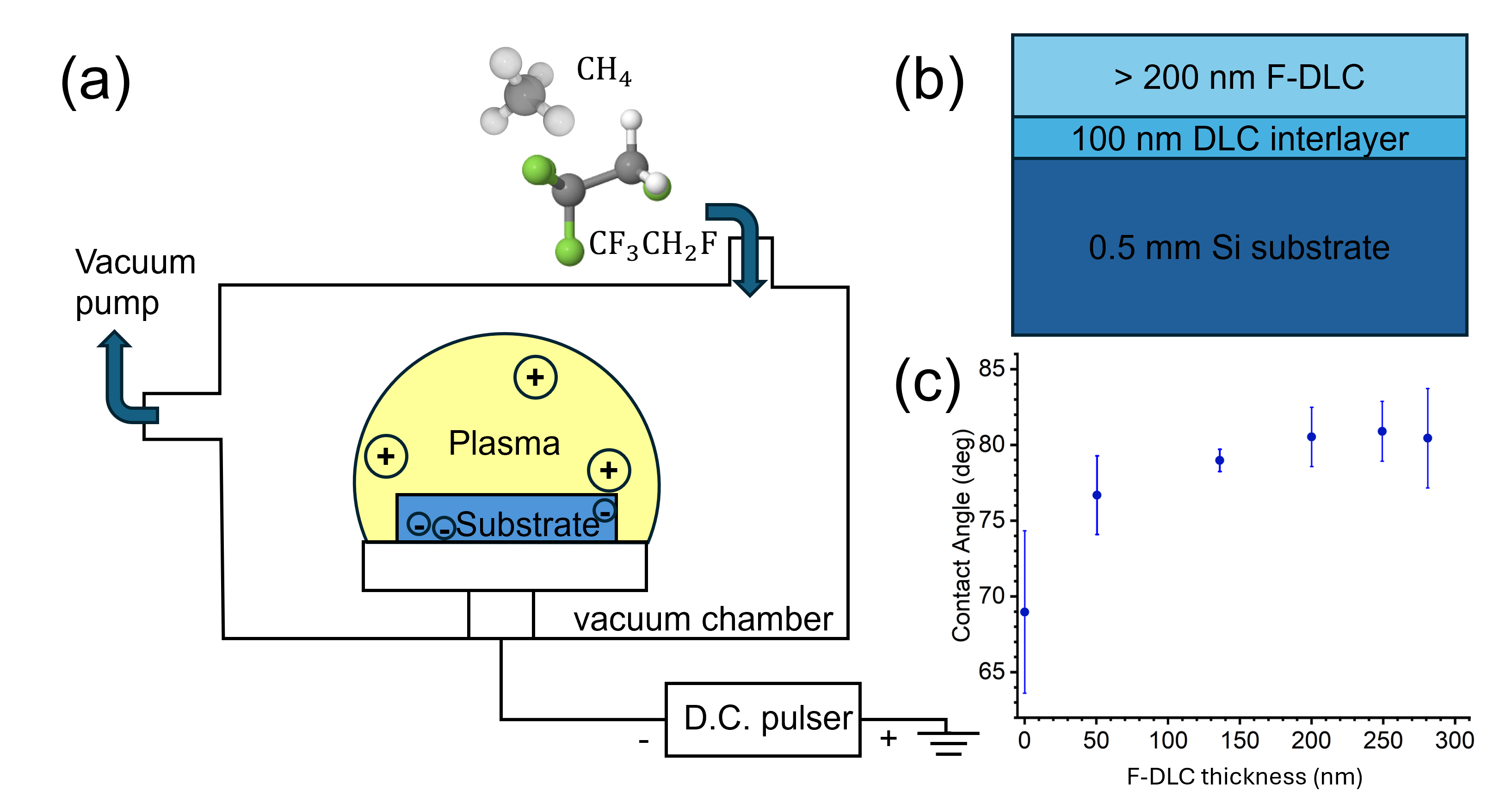}
        \label{fig:pecvd}
    \end{subfigure}

    \caption{(a) PIII-PECVD setup for depositing F-DLC coatings. (b) Illustration of the cross-section of the F-DLC samples studied in this work. (c) Measured water contact angle as a function of deposited F-DLC thickness (F:C=0). All samples from different F concentrations reported in the rest of the paper exceed the critical thicknesses (around 200 nm) such that the contact angle is not thickness-dependent.}
    \label{fig:pecvd_setup}
\end{figure}

All F-DLC films used in this work were deposited onto single crystalline silicon substrates using the PIII-PECVD setup shown in Figure \ref{fig:pecvd_setup} (a) \cite{muley2022optimizing} explained in Section \ref{Preparation for F-DLC samples}. To clean substrates prior to deposition, Ar sputtering was performed at an Ar flow rate of 100 sccm and a pressure of 18 mTorr for 20 minutes. The Direct current (D.C.) pulser parameters were set to 5.0 kV, 5.0 kHz, and 10.0 $\mu$s for Ar sputtering.

\begin{table}[ht]
\centering
\begin{tabular}{|p{3.1cm}||p{0.5cm}|p{0.5cm}|p{0.5cm}|p{0.8cm}|p{0.5cm}|p{0.5cm}|p{0.8cm}|p{1cm}|p{0.8cm}|}
 \hline
 CH$_4$ (sccm) & 50 & 80 & 60 & 40 & 20 & 10 & 5 & 2.5 & 0 \\
 \hline
 CF$_3$CH$_2$F (sccm) & 0 & 10 & 10 & 10 & 10 & 15 & 14.15 & 15.425 & 14.49 \\
 \hline
 F:C & 0.0 & 0.4 & 0.5 & 0.667 & 1.0 & 1.5 & 1.7 & 1.85 & 2.0 \\
 \hline
\end{tabular}
\caption{Gas compositions used for the deposition of F-DLC films via PECVD.} 
\label{table:gas}
\end{table}

During F-DLC deposition, a pressure of 0.15 Torr and a DC pulse of 2.5 kV, 5.0 kHz, and 10.0 $\mu$s were used. The gas composition comprising of CH$_4$ and CF$_3$CH$_2$F was varied to obtain F-DLC films with different F:C ratios, as shown in Table \ref{table:gas}. 

Due to the low adhesion of F-DLC thin films to Si substrates, a 100-nm hydrogenated DLC interlayer was deposited using CH$_4$ first to improve adhesion \cite{yao2004structural}. A detailed discussion of the effect of the hydrogenated DLC interlayer and mechanical properties of F-DLC films with varied fluorine content is shown in Section \ref{mechanical results}. In F-DLC, the fluorine-carbon bonds are chemically inert and do not easily form interfacial bonds with Si, and the weak adhesion of Si/F-DLC makes the films prone to delamination. The addition of interlayer DLC improves the bond strength with Si-C, enhancing adhesion. The carbon in DLC and carbon-fluorine bonds in F-DLC will also have a stronger interaction than Si-F interaction, leading to an enhanced adhesion. All F-DLC thin films reported here have a thickness of at least 200 nm (Figure \ref{fig:pecvd_setup} (b)). Maintaining a minimum F-DLC thickness is crucial for a consistent contact angle measurement, since thicker films exhibit reduced porosity and increased structural uniformity. We evaluated the sensitivity of the measured contact angles to film thickness in Figure \ref{fig:pecvd_setup} (c) and observed that the contact angles stabilize for thicknesses greater than 200 nm for all fluorine concentrations.

\section{Compositional analysis of F-DLC films}

Using X-ray photoelectron spectroscopy (XPS) described in Section \ref{XPS Analysis 1}, we analyzed the presence of C (from the carbon matrix), F (from fluorination), O (from oxidation), and Si (from the silicon substrate) in the F-DLC samples. The atomic percentages were calculated using the survey scan, specifically: C\% from C1s peak, F\% from F1s and F2s peaks, and O\% from O1s and O2s peaks.

\begin{figure}[hbt!]
    \centering
        \includegraphics[width=\textwidth]{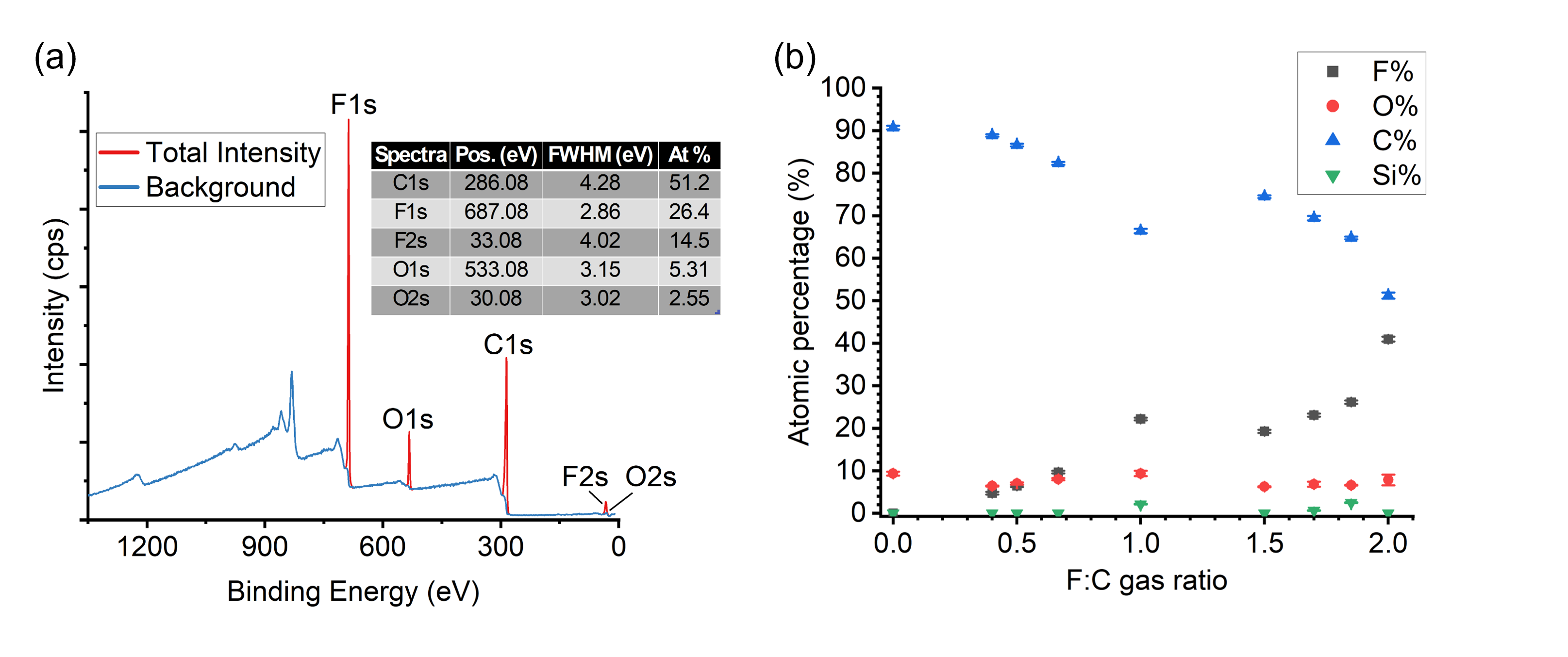}
        \label{fig:xpssurvey}
    \caption{(a) XPS survey scan of F:C=2.0 F-DLC sample. (b) XPS survey scan result with C\%, F\%, O\%, Si\% (atomic percentage in the deposit) corresponding to the F:C (atomic ratio in process gas mixture).}
    \label{fig:survey_scan_data}
\end{figure}

Figure \ref{fig:survey_scan_data} (a) displays the survey scan for the highest F:C ratio of 2, highlighting the elemental compositions observed. As shown in Figure \ref{fig:survey_scan_data} (b), as the F:C ratio increases, we observe an increase in F\% and a corresponding decrease in C\%. The outlier at F:C = 1 could be attributed to localized variations in the deposition process, such as fluctuations in gas flow rates or plasma conditions, leading to inconsistent incorporation of fluorine in the F-DLC film at that specific ratio. 

Across all F-DLC samples, the O\% remains between 5\% and 10\% (Figure \ref{fig:survey_scan_data} (b)) with no strong correlation with the fluorine level in the gas flow. Meanwhile, the Si\% values are all below 4\%, suggesting minimal contribution from the substrate during XPS analysis. For consistency, we will use F\% instead of F:C ratio in the rest of the paper to represent the fluorine content in the F-DLC deposits.

We performed narrow scans of the C1s peak to analyze the carbon bonding states, as explained in Section \ref{XPS Analysis 1}. Given the large number of possible chemical bonds that can be present in F-DLC films, a consistent approach to devolution and fitting of the C1s peak is crucial. Here we have ensured that our observed peak positions for the sp$^2$ carbon-carbon, sp$^3$ carbon-carbon, carbon-oxygen (C-O, C=O), and carbon-fluorine (C-CF, C-F, C-F\(_2\)) functional groups are well-aligned with those reported in prior literature \cite{ahmed2012evaluation, tressaud_nature_1996} and implemented constraints such that the fitted lineshapes and linewidths are consistent for each functional group across all of our samples. Further details are provided in Section \ref{XPS Analysis 1}. Additionally, a detailed table of all chemical groups identified with different F \% is included in Supplemental Material (Table S2).

\begin{figure}[hbt!]
    \begin{subfigure}[b]{1\textwidth}
        \centering
        \includegraphics[width=\textwidth]{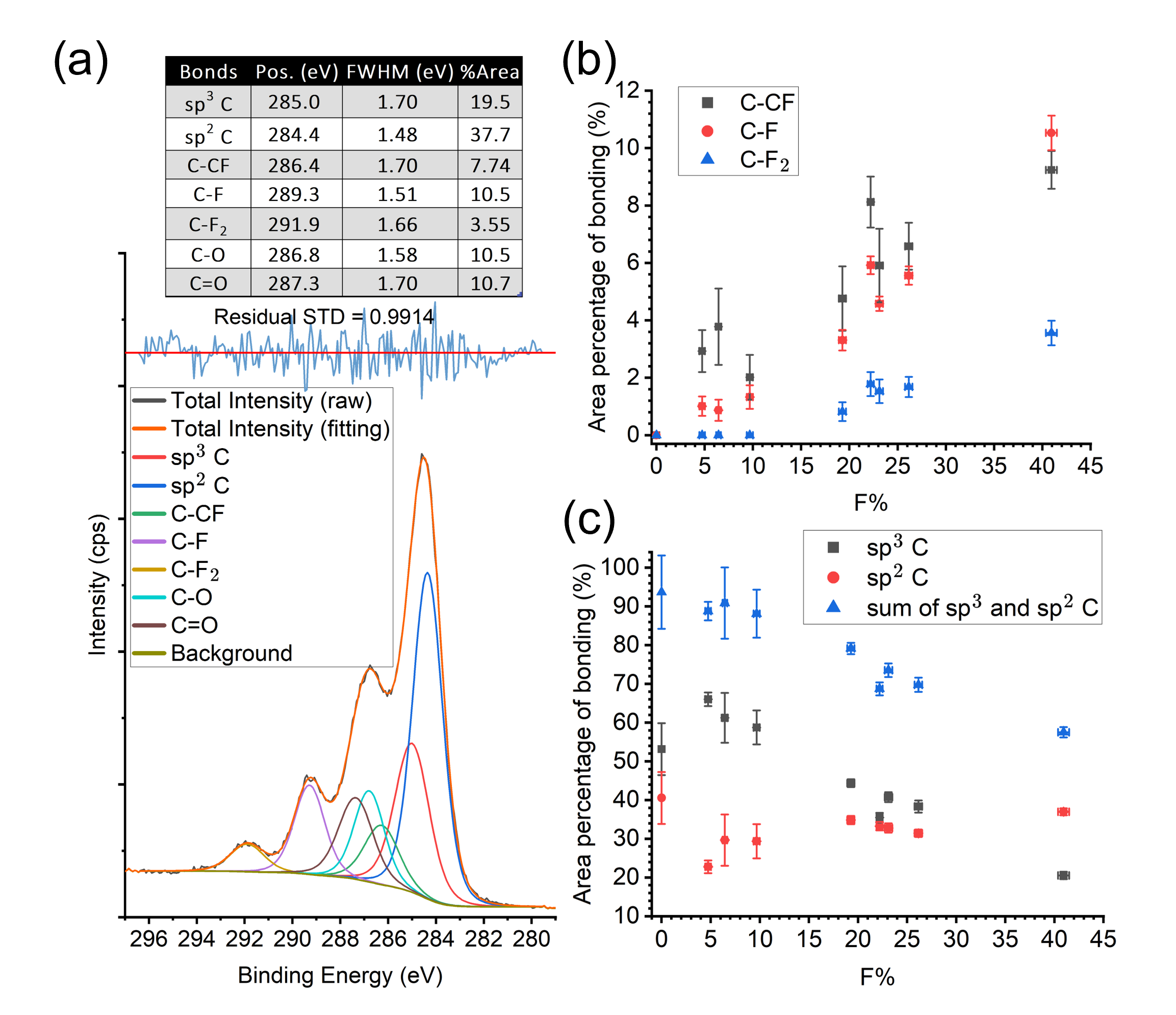}
        \label{fig:xps_deconvolution_c_c}
    \end{subfigure}
    
    \caption{(a) XPS C1s peak deconvolution of the F\% = 40.9\% F-DLC sample displaying showing various chemical bonds including sp$^3$ C, sp$^2$ C, C-CF, C-F, C-F$_2$, C-O, and C=O. (b) Plot of the area percentage of carbon-fluorine, (c) sp$^3$ and sp$^2$ carbon bonding as a function of the fluorine atomic percentage in the deposit, illustrating how the surface chemistry changes with increasing fluorine content. The error bars in y-axis direction of plots (b,c) represent the uncertainty in the area percentage of bonding, calculated using Monte Carlo simulations in CasaXPS, taking into account the uncertainties in the peak positions, areas, and the full-width-half-maximum (FWHM) of each bond.}
    \label{fig:xps_data}
\end{figure}

Figure \ref{fig:xps_data} (a) provides a representative C1s XPS spectrum and set of deconvolution and fitting results (for F\% = 40.9\%). As the fluorine content increases, there is a corresponding rise in the C-CF and C-F percentages, along with the appearance of C-F\(_2\) surface groups beyond an F\% of 20\% (Figure \ref{fig:xps_data} (b)). These observations are in agreement with previous studies using other process gases \cite{su2010modification, bendavid2009properties}, suggesting that the use of 1,1,1,2-tetrafluoroethane effectively creates the desired carbon-fluorine bonding. Additionally, Figure \ref{fig:xps_data} (c) shows a decrease in the sp\(^3\) C percentage and a corresponding slight increase in the sp\(^2\) C percentage in F-DLC films. As will be seen in Sections 4 and 5, the increase in the sp\(^2\) C / sp\(^3\) C ratio (Supplemental Material, Figure S3) correlates with a change in physical properties \cite{paprocki_comparative_2019}, specifically a reduction in wettability (observed in hydrogenated DLC \cite{lee_correlation_2007}) and an increase in surface roughness (in F-DLC films from other deposition methods\cite{ahmed2012evaluation, jiang2013effect}).


\section{Wettability and surface free energy of F-DLC films}

\begin{figure}[hbt!]
     \begin{subfigure}[b]{1\textwidth}
         \centering
         \includegraphics[width=\textwidth]{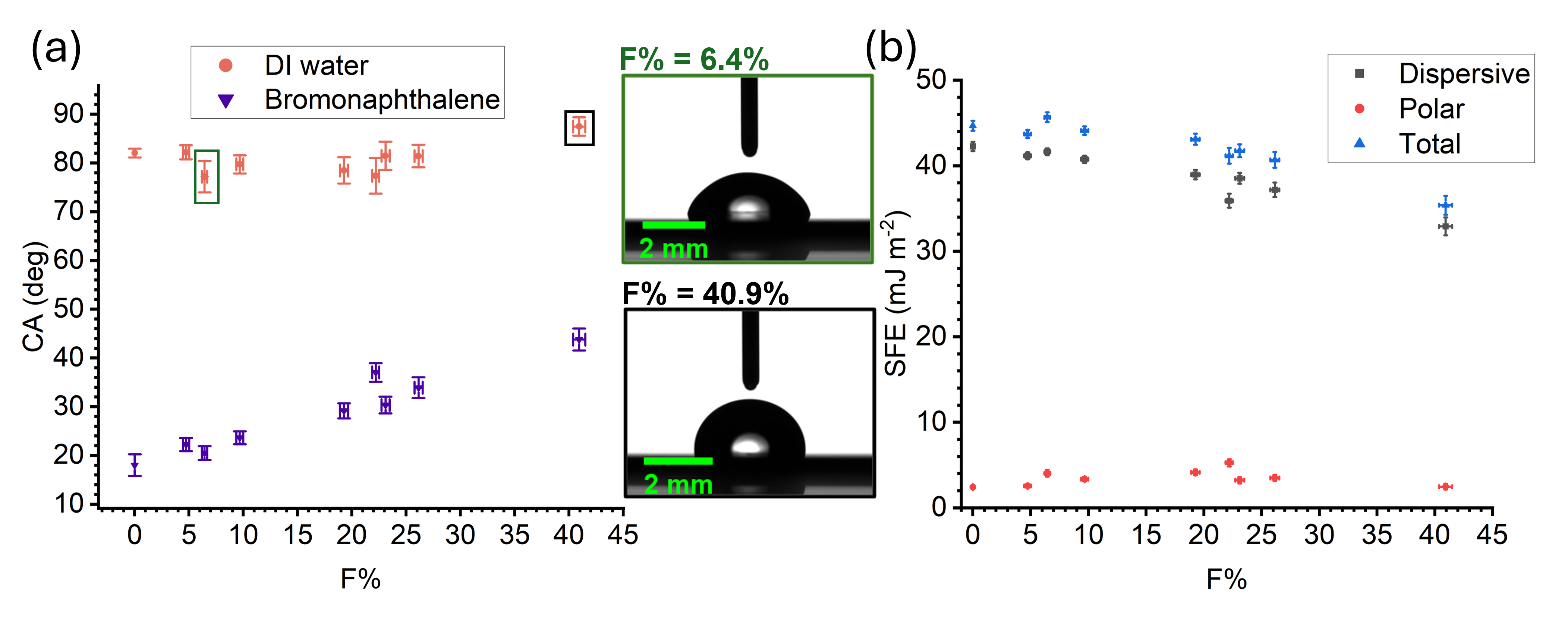}
         \label{fig:surface energy result}
     \end{subfigure}
     \hfill
    \caption{(a) Contact angle (CA) results of DI water and bromonaphthalene on F-DLC with various F\%. In particular, the images of F\% = 6.4\% and 40.9\% DI water's contact angle are shown. (b) Surface free energy is calculated based on contact angle measurements for F-DLC films with different F\%.}
    \label{fig:SFE and CA data}
\end{figure}

The surface wettability of the F-DLC films was evaluated using optical contact angle measurements (sessile drop method) with a polar liquid (DI water) and a non-polar liquid (bromonaphthalene), as detailed in Section \ref{Optical Contact Angle}. We observed that high F contents (F\% above 25\%) increase the contact angles (Figure \ref{fig:SFE and CA data}(a)) and thus reduce the wettability of the F-DLC films in both polar and non-polar liquids. This effect is more prominent for the latter, with the bromonaphthalene contact angle doubling from 20° to 40° with the incorporation of fluorine.

The DI water contact angle shows a more intricate correlation with the F\%. The DI contact angle first reduces at low F\% values and then starts to increase once the F\% exceeds past 20\%. The onset of contact angle increase is correlated with the increased presence of C-F\(_x\) bonds and decreased carbon-carbon bonding as seen in Figure \ref{fig:xps_data} (b) and (c). This observation is consistent with those reported in \cite{bendavid2009properties, bendavid2010properties, kasai1986surface, zhang2020effect}. The presence of C-F\(_2\), in particular, has been attributed to disrupting the aromatic ring symmetry \cite{jiang2006influence} (hence the reduced carbon-carbon bonding) and the transition from diamond-like structure to polymer-like structure \cite{zhang2015recent}.

To further analyze the surface properties of F-DLC films, we calculated the \(\sigma_s^p\) and \(\sigma_s^d\) components of the surface free energy using the method described in Section \ref{Surface Free Energy calculation}. These calculations are based on contact angle measurements of DI water and bromonaphthalene. The results of the \(\sigma_s^p\), \(\sigma_s^d\), and total surface free energy at fluorine contents ranging from 6.4\% to 40.9\% in F-DLC films are shown in Figure \ref{fig:SFE and CA data} (b). In the figure, \(\sigma_s^p\) exhibits minor fluctuations corresponding to the DI water contact angle behavior. Conversely, the \(\sigma_s^d\) component consistently decreases as the fluorine content increases, correlating with the rise in bromonaphthalene contact angle. Since the dispersive component is dominant over the polar component, the total surface free energy is primarily influenced by the larger decrease in the dispersive component. The total surface free energy decreases by approximately 10 mJ/m$^2$ with increasing fluorine content in F-DLC films from 6.4\% to 40.9\%, consistent with the findings of Ishihara et al. \cite{su2010modification, ishihara2006antibacterial}.

The monotonic increase in the bromonaphthalene contact angle can be attributed to surface chemistry changes due to fluorine incorporation. Fluorine is highly electronegative and has low polarizability. As fluorine content increases in the film, more C-F bonds replace C-C and C-H bonds, which reduces London dispersion forces and lowers the dispersive component of the surface energy \cite{lemal2004perspective}. Since bromonaphthalene interacts predominantly through dispersive forces, the reduced dispersive surface energy on F-DLC leads to weaker interactions with the liquid, resulting in a higher contact angle. The observed decrease in surface energy with increasing fluorination is in agreement with theoretical expectations: fluorine incorporation reduces the dispersive component of surface energy due to its low polarizability, which also aligns with prior studies on other non-polar liquids contact angle \cite{su2010modification, ishihara2006antibacterial}.

\section{Correlation of wettability with surface morphology}

\begin{figure}[ht]
      \centering
        \includegraphics[width=0.8\textwidth]{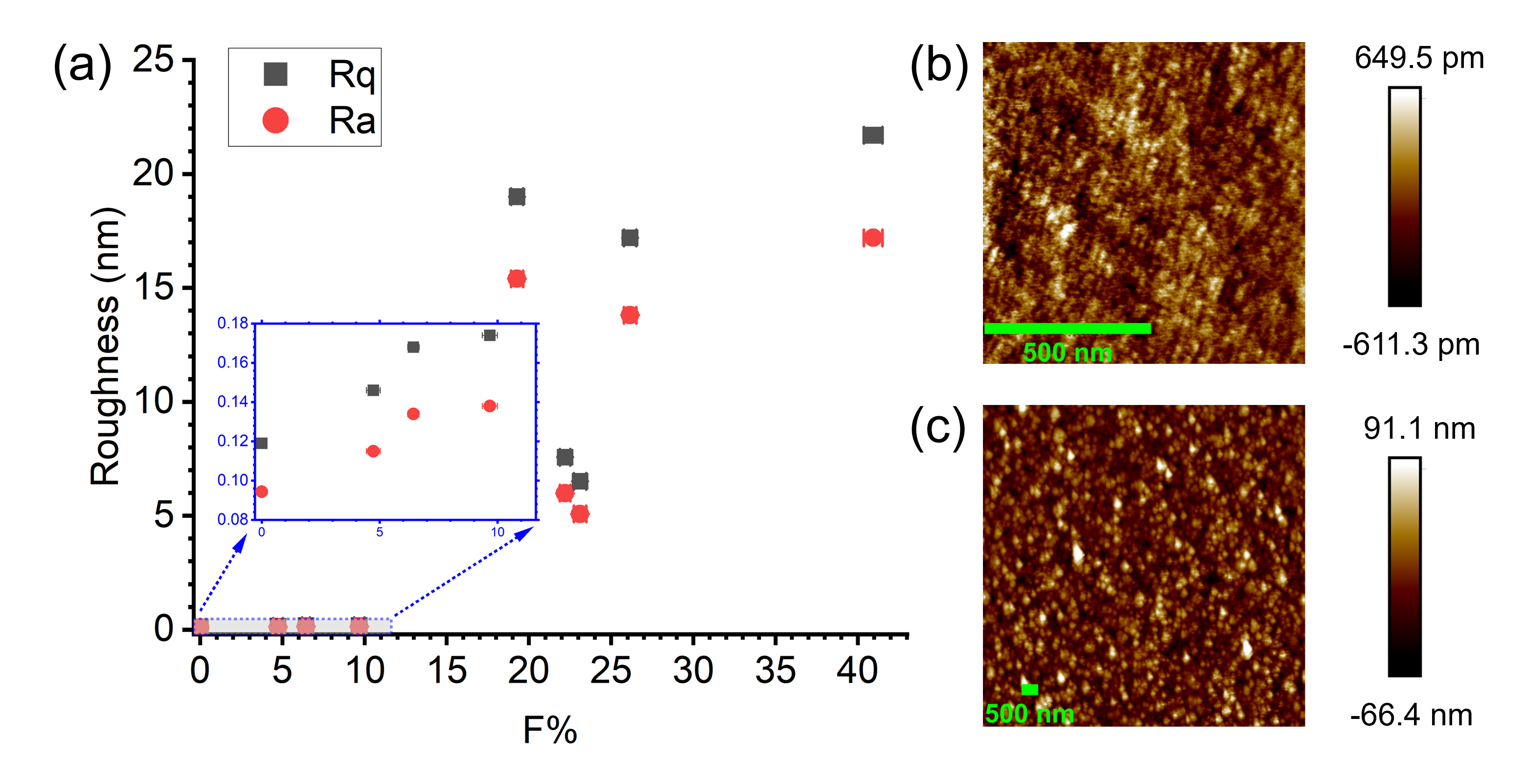}
         \label{fig:afm_with pic}
     \hfill
    \caption{(a) AFM $R_a$ and $R_q$ for different F\%. (b) and (c) AFM images of a 6.4 F\% and 40.9 F\% are shown. }
    \label{fig:afm data}
\end{figure}
Generally, a higher fluorine content enhances the hydrophobicity of F-DLC surfaces due to the chemical inertness provided by C-F\(_x\) bonds \cite{zhang2015recent}. However, our observations of DI contact angle in Figure \ref{fig:SFE and CA data}(a) suggest an opposing factor that prevents the DI water contact angle from exceeding 90°, even with increased C-F\(_x\) at higher fluorine levels. The likely reason is the effect of surface roughness \cite{yao2004structural}, which is corroborated by our atomic force microscopy (AFM) results (Figure \ref{fig:afm data}, method described in Section \ref{AFM method}). 

The surface roughness of our F-DLC films, represented by root mean square roughness (Rq) and average roughness (Ra) values, increases significantly with higher fluorine content. Specifically, Rq rises from $\sim$1 nm to 10-20 nm for F\% $>$ 20\%, which is consistent with previous observations reported for F-DLC surfaces \cite{su2010modification, ahmed2012evaluation, zhang2020effect}, although during the transition around F\%$\approx$20\% the roughness appeared to fluctuate. Additionally, AFM images of the F-DLC surfaces at 6.4 \% and 40.9 \% are demonstrated in Figure \ref{fig:afm data} (b) and (c) revealing the presence of small granular features at higher F \% (a characteristic also reported in F-DLC coatings by Jiang et al. \cite{jiang2013effect}), and an SEM topview image of a low F\% and a high F\% is shown in Supplemental Material (Figure S4).

To examine the impact of surface roughness on the wettability of F-DLC surfaces with DI water, we applied Wenzel's equation \cite{wenzel_resistance_1936}, which relates the contact angle on a rough surface to that on a smooth surface of the same material. This equation is valid since our drop size is much larger than the surface roughness \cite{wolansky1999apparent, marmur_when_2009}. Wenzel's equation predicts that rougher surfaces decrease the contact angle of DI water when the intrinsic angle (smooth surface) < 90\textdegree. This explains why, as shown in Figure \ref{fig:SFE and CA data} (a), the DI contact angle does not increase as expected despite higher fluorine content and more C-F\(_x\) surface groups. 

Similarly, increasing roughness at higher F content reduces the bromonaphthalene contact angle, as its intrinsic contact angle is < 90\textdegree. The results are consistent with Wenzel's prediction, but for bromonaphthalene, the effect of roughness on the contact angle is less pronounced. The significant reduction in dispersive surface energy at higher fluorine content leads to a more prominent change in bromonaphthalene contact angle, which is more influenced by the dispersive forces. Thus, the reduction in dispersive surface energy dominates the wettability behavior for bromonaphthalene, and roughness does not fully counteract this effect. Notably, when the intrinsic contact angle is > 90\textdegree, roughness can amplify hydrophobicity, leading to superhydrophobic surfaces \cite{terriza2012roughness}.

\section{Mechanical properties of F-DLC films}
\label{mechanical results}

To further examine the relationship between fluorine concentration and mechanical properties of the F-DLC films, we performed nanoindentation to measure the hardness ($\mathrm{H}$) and reduced elastic modulus ($\mathrm{E_r}$), and nanoscratch to determine the critical load ($L_c$), a key indicator of adhesion between the substrate and the coating \cite{beake2006influence}. Experimental details for these tests are provided in Section \ref{mechanical method}.

\begin{figure}[ht!]
      \centering

        \includegraphics[width=1\textwidth]{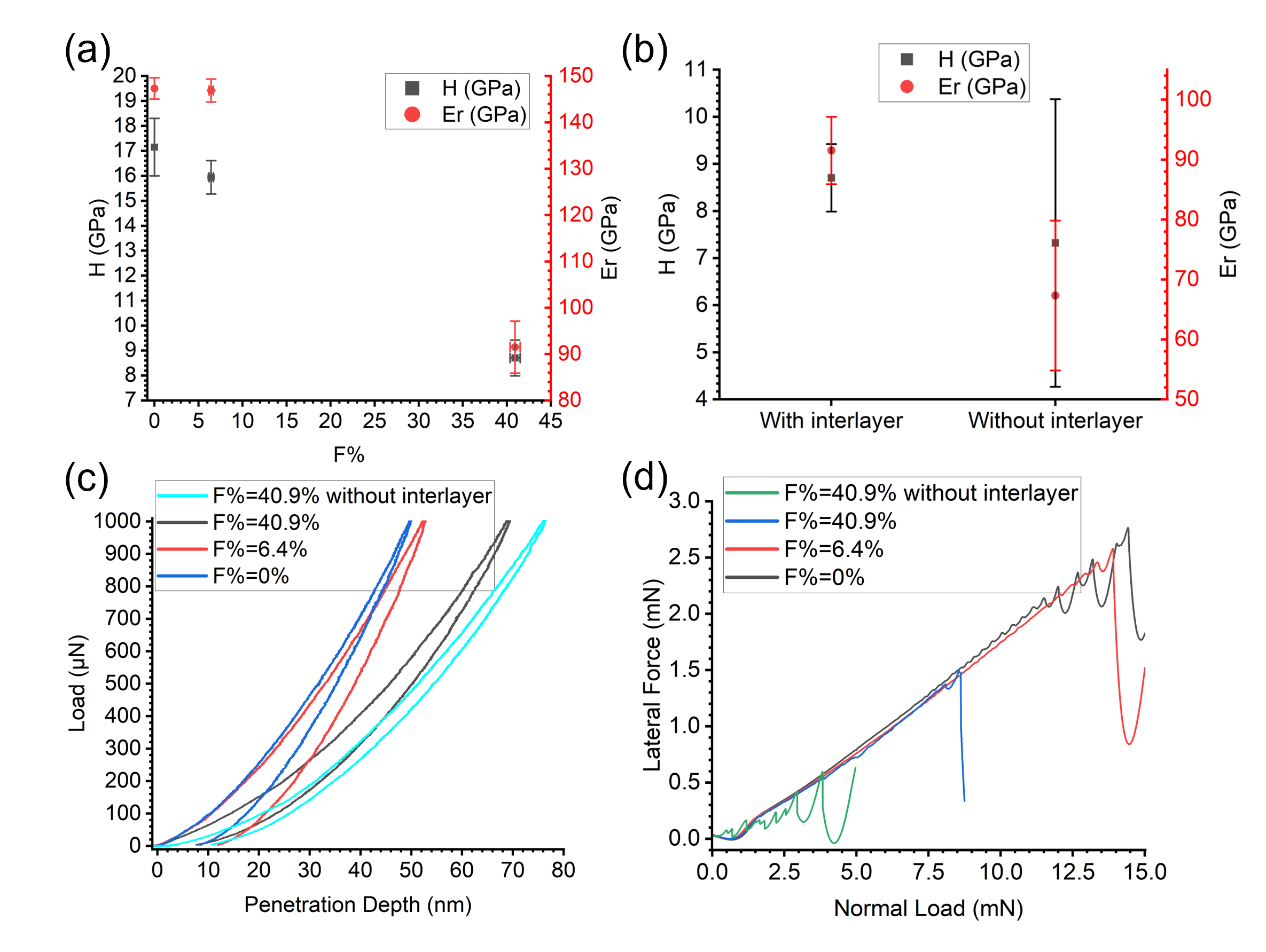}
         \label{fig:mechanical}
     \hfill
    \caption{(a) Hardness ($\mathrm{H}$) and reduced elastic modulus ($\mathrm{E_r}$) of F-DLC samples with F\%=0\%, F\%=6.4\%, and F\%=40.9\%, all with hydrogenated DLC interlayers. (b) $\mathrm{H}$ and $\mathrm{E_r}$ of F\%=40.9\% samples, with and without interlayers. (c) representative load/unload curves showing penetration depth variations during nanoindentation tests under different conditions of F-DLC films. (d) representative lateral force vs. normal load from nanoscratch tests for different conditions of F-DLC films.}
    \label{fig:mechanical data}
\end{figure}

Consistent with prior literature \cite{yao2004structural} that points to increased mechanical flexibility of F-DLC, our F-DLC films show a lower $\mathrm{H}$ and $\mathrm{E_r}$ with increasing F content (Figure \ref{fig:mechanical data} (a), evidenced by the higher penetration depth and shallower unload curve in Figure \ref{fig:mechanical data} (c). The increased mechanical flexibility can be explained by our XPS analysis which shows higher sp$^2$ and lower sp$^3$ carbon as F\% increases, since fluorine incorporation disrupts the sp$^3$ network and promotes the mechanically weaker sp$^2$ structures. The adhesion of high F\% F-DLC films is significantly weaker than that of low F\% films, as shown by a lower critical load of total film failure (Figure \ref{fig:mechanical data} (d)).

The hydrogenated DLC interlayer enhanced the adhesion at high F\% significantly (Figure \ref{fig:mechanical data} (d)). The film without an interlayer exhibited cracking almost immediately upon scratching and failed completely at a load below 3 mN. On the other hand, the film with interlayer had significantly less cracking, with failure occuring near the load of 8 mN. Additionally, the mean $\mathrm{H}$ and $\mathrm{E_r}$ are slightly higher for the F\%=40.9\% film with an interlayer compared to the film without an interlayer (Figure \ref{fig:mechanical data} (b)). This difference can be due to the poor adhesion in the sample without an interlayer, where poor adhesion allows lateral displacement of the coating during indentation, leading to increased penetration depth (lower hardness). In contrast, samples with interlayer have better adhesion and structural support to the F-DLC film, reducing lateral deformation, resulting in a lower penetration depth and better recovery during unloading. The higher standard deviation for F-DLC without interlayer indicates heterogeneity in the film.

\section{Conclusion}
\label{Conclusion}

This study provides a comprehensive analysis of F-DLC films synthesized using PIII-PECVD, with F$_3$CH$_2$F as the fluorine precursor. This is a scalable and environmentally safe process that, to the best of our knowledge, had yet to be reported and would be particularly well-suited for the deposition of low-surface-energy DLC films on complex substrates.


Our analysis integrates findings from contact angle measurements, total surface free energy calculations, XPS survey and C1s peak scans, AFM data, and mechanical tests. We show that F-DLC coatings formed with CF$_3$CH$_2$F exhibit properties comparable to those produced with other fluorinated gases such as CF\(_4\) and C\(_2\)F\(_4\) \cite{su2010modification, grischke1995application, jiang2013effect}. The introduction of CF$_3$CH$_2$F increased the fluorine content up to 40\%, leading to the formation of  C-F\(_x\) surface groups, which resulted in an overall increase in the contact angle for both polar and non-polar liquids and a reduction in the total surface energy driven by the decrease in the dispersive forces. The observed reduction in surface energy (around 10 $mJ/m^2$) is sufficient in improving anti-bacterial preoprties of the films, based on prior studies \cite{ishihara2006antibacterial}. Our work also highlights the role of surface roughness as a counteracting factor that limits the increase in DI water contact angle. 

Additionally, we demonstrate that F-DLC films with high F\% have significantly lower hardness and reduced elastic modulus in comparison to DLC films, which is consistent with increased mechanical flexibility in F-DLC. Flourine incorporation at high F\% also tends to reduce adhesion, which is somewhat mitigated by the use of a DLC interlayer between the substrate and the film. Future characterizations with Raman spectroscopy may provide further insights on the internal stress and disorder of F-DLC films. Finally, the long-term hydrophobic properties of the F-DLC film are important for its industrial applications. Changes in surface energy, especially related to oxygen groups, affect its performance over time. This is currently being studied and is crucial for ensuring its effectiveness in real-world use.


\section{Materials and Methods}
\label{Methods and Results}

\subsection{Preparation of F-DLC samples}
\label{Preparation for F-DLC samples}

The single crystalline silicon substrates used in this work are mechanical grade, single-side polished, and have a thickness of 0.5 mm. These substates are cut into $\sim$ 15 mm $\times$ 15 mm squares and solvent cleaned (involving sonication in acetone for 10 minutes, rinsing with methanol and isopropyl alcohol separately, and then drying with $N_2$ gas).

The PIII-PECVD setup described in Section \ref{growth} has a base pressure of 2 mTorr and utilizes a pulser from Applied Energetics, U.S.A. Chemours Freon™ 134a refrigerant (1,1,1,2-tetrafluoroethane) was obtained in the form of liquefied compressed gas and used without further purification.

The thickness measurements are performed using Filmetrics F20 optical thin-film measurement system and confirmed with spectroscopic ellipsometry (J. A. Woollam V-VASE). Representative ellipsometry data and the optical constants of the films can be found in the Supplemental Material section 1.

\subsection{XPS Analysis: Elemental Composition and Surface Group Quantification}
\label{XPS Analysis 1}

We used X-ray photoelectron spectroscopy (XPS) \cite{watts2019introduction}, performed using a Thermo K Alpha X-ray Photoelectron Spectrometer, to measure the atomic composition of F-DLC sample surfaces at various F\% levels, aiming to understand how the presence of different elements and surface groups changes with varying fluorine content. We selected one random region ($\sim$ 200 $\mu$m in diameter) from one sample per condition for XPS analysis. 

Initially, a survey scan was conducted across the entire range of binding energies, calculating elemental percentages based on peak areas, background signals, and relative sensitivity factors. To acquire the F\%, survey scan data were collected from a binding energy range of 10 eV to 1350 eV, with a step size of 1 eV, pass energy of 200 eV (indicating the kinetic energy of the electrons as they pass through the analyzer, balancing resolution and signal intensity), and a dwell time of 50 ms.

To gain deeper insights into the surface groups influencing contact angles that will be elaborated later in this paper, we deconvoluted the C1s peak into distinct chemical bonds. The C1s scan was performed over a range of 279 eV to 298 eV, with an energy step size of 0.08 eV, pass energy of 20 eV, 15 scans, and a dwell time of 100 ms, ensuring that the noise level did not impede the deconvolution process.

We fitted the curves of C1s scan and constrained full-width-at-half-maximum (FWHM) values ranging from 1.0 eV to 1.7 eV.  Although these peak positions tend to shift in different samples due to their charging effect, we centered the main C1s peak to 284.6 eV \cite{su2010modification} to compensate the differences in surface charging and calibrate peak positions.

For F-DLC samples at each F\%, we found sp\(^2\) C at 284.4 eV, sp\(^3\) C at 285.0 eV, C-CF at 286.4 eV, C-F at 289.3 eV, and C-F$_2$ at 291.9 eV \cite{ahmed2012evaluation, tressaud_nature_1996}. According to Filik's \cite{filik_xpsand_2003} and Tai's work \cite{tai_correlation_2006}, fitting the XPS curve with CO bonds for hydrogenated DLC samples results in a closer fit. C-O and C=O bonds were found at 286.8 eV and 287.3 eV, respectively \cite{ahmed2012evaluation, wei_effect_2020}. Moreover, the contribution from the C-CF\(_2\) bond was negligible, so it was not included. These findings highlight the reliability and accuracy of our fitting, as they are consistent with previous studies while providing accurate FWHM values and peak positions. In this study, instead of using a baseline correction to eliminate the background signal, we applied a Shirley background to keep the raw data unmodified for better comparability with other studies and account for the nonlinear background signal while avoiding overcorrecting the spectra. Additionally, after optimizing the peak shapes collectively on CasaXPS, we found that using a Lorentzian-based lineshape "LA(2,100)" provided the best fit with the least variation from the actual data.

\subsection{Optical Contact Angle}
\label{Optical Contact Angle}

We optically measured the contact angles of liquids on F-DLC samples with varying F\% using sessile drop method. Measurements can be influenced by factors such as ambient temperature, humidity, sessile drop volume, surface cleaning, and time after deposition \cite{schuster2015influence, zhang_effects_2022}. Even under controlled conditions, the results can vary by a few degrees. Therefore, we measured the contact angle at one random point on each of the four samples per F\% to improve statistical reliability.

In this paper, the DI Water used as a liquid phase is ASTM Type I (ultrapure) water produced by the Barnstead EASYpure II UV water purification system. The bromonaphthalene is obtained from Thermo Fisher Scientific (97\%) and used without further purification. The measurements were taken at room temperature, with DataPhysics OCA15EC. In each measurement, an 8 \(\mu L\) liquid droplet was dispensed at the center of each sample at a rate of 0.5 \(\mu L/s\), while a camera recorded the process in 30 frames per second. The contact angle was measured in each frame and once it reached a steady state, the contact angle image was taken and the value is calculated from elliptical fitting.

\subsection{Surface free energy calculation}
\label{Surface Free Energy calculation}

Calculating total surface free energy in addition to measuring contact angles provides a more comprehensive understanding of the wettability and interfacial properties of F-DLC samples. To quantify the surface energy, we employed the Owen and Wendt \cite{owens1969estimation} method which allow us to distinguish between the dispersive and polar components of surface interactions with different types of liquids. 

Based on Fowkes' theory \cite{fowkes_attractive_1964}, we calculated the dispersive (\(\sigma^d\)), polar (\(\sigma^p\)), and total surface free energy (\(\sigma\)) of F-DLC samples at each F\%. These components follow the relationship:

\begin{equation}\label{eq:1}
    \sigma_{s/l} = \sigma_{s/l}^d + \sigma_{s/l}^p.
\end{equation}

Here, \(s\) stands for solid and \(l\) stands for liquid. The surface free energy calculations are based on the known \(\sigma_l^d\) and \(\sigma_l^p\) components of the liquids used. We aim to determine the \(\sigma_s^d\) and \(\sigma_s^p\) components of the F-DLC films as a function of F\%.

The selection of liquids is important for the precision of surface free energy calculations, and using a combination of polar and non-polar liquids yields the most accurate results with the smallest uncertainty. For simplicity, we chose deionized (DI) water ($\sigma_l^d = 21.8 \, \mathrm{mJ/m^2}$, $\sigma_l^p = 51.0 \, \mathrm{mJ/m^2}$) as the polar liquid and bromonaphthalene ($\sigma_l^d = 44.4 \, \mathrm{mJ/m^2}$, $\sigma_l^p = 0.0 \, \mathrm{mJ/m^2}$ ) as the nonpolar liquid \cite{fialho2018stainless}.

According to Fowkes' theory for the work of adhesion (\(W_{sl}\))\cite{owens1969estimation}, \(W_{sl}\)  is given by:

\begin{equation}\label{eq:2}
    W_{sl} = 2 \left( \sqrt{\sigma_l^d \sigma_s^d} + \sqrt{\sigma_l^p \sigma_s^p} \right),
\end{equation}

Then, substituting this expression for the work of adhesion into Young-Dupré's equation:

\begin{equation}\label{eq:3}
    W_{sl} = \sigma_l (1 + \cos \theta),
\end{equation}

where \(\theta\) is the contact angle of the liquid on the solid surface, we obtain the following relationship \cite{janczuk1989surface}:

\begin{equation}\label{eq:fowkes}
    \sqrt{\sigma_l^d \sigma_s^d} + \sqrt{\sigma_l^p \sigma_s^p} = \frac{\sigma_l (1 + \cos \theta)}{2}.
\end{equation}

To analyze the surface energy components (\(\sigma_s^d\) and \(\sigma_s^p\)) of F-DLC coatings at different F\%, we used Equation \ref{eq:fowkes} twice. First, we utilized the surface energy values and contact angle measurements of bromonaphthalene to calculate \(\sigma_s^d\) for the coatings. Then, we used the surface energy and contact angle values for DI water to determine \(\sigma_s^p\).

Since bromonaphthalene has \(\sigma_l^p = 0\), \(\sigma_l = \sigma_l^d\), we can simplify the equation as follows:

\begin{equation}\label{eq:sigmasd}
    \sigma_s^d = \frac{\sigma_l^d (1 + \cos \theta)^2}{4}.
\end{equation}

Here, \(\theta\) is the contact angle with bromonaphthalene. Using this equation, we calculated \(\sigma_s^d\) for F-DLC as a function of fluorine content. 

Next, we performed contact angle measurements with DI water to obtain \(\sigma_s^p\). Using the measured contact angles, \(\sigma_l^p\) and \(\sigma_l^d\) of DI water, and \(\sigma_s^d\) from the previous step, we calculated \(\sigma_s^p\) of the films at each F\%, using Equation \ref{eq:fowkes}.

The error in surface energies was derived from the uncertainties in the measured contact angles of the two liquids, using error propagation to account for these errors in the final analysis.

\subsection{Atomic Force Microscopy}
\label{AFM method}
To quantify the roughness of the deposited F-DLC films with varying F\% and to verify their correlation with surface wettability, we performed AFM on the samples. The Bruker Dimension Icon AFM, equipped with RTESPA-300-125 tips, was used in tapping mode in air. We selected 3 random areas of $1 \mu m \times 1 \mu m$ on one sample and measured their roughness, comparing these measurements across samples with the same F\%, which showed that evaluating one region per F\% condition was sufficient to assess surface roughness.

The AFM scans were conducted at a scan rate of 1 Hz with a resolution of 256 x 256 pixels. The drive amplitude (the amplitude of the oscillating cantilever is driven), set point (the value of the cantilever oscillation amplitude from the feedback loop), and integral and proportional gains (sensitivity of the feedback loop to the tip's vibration \cite{Step-by-step_AFM}) were adjusted to ensure optimal alignment of the trace and retrace lines.

\subsection{Nanoindentation and Nanoscratch}
\label{mechanical method}
Nanoindentation and nanoscratch experiments were performed on a Hysitron Ti 950 Triboindenter (calibrated with a standard fused quartz sample) using a Berkovich (Ti-0039) diamond tip. Data acquisition and analysis were conducted using the Triboscan software.

For the nanoindentation test, the load/unload data were shifted horizontally to zero point before fitting. The unload curve fitting procedure and the calculation of hardness ($\mathrm{H}$) and reduced elastic modulus ($\mathrm{E_r}$) can be found in the study by Muley et al. (\cite{muley2022optimizing}). From our measurement results, residual depth to maximum depth ratio were restricted to less than 0.7, indicating minimal pile-up effect that could influence the accuracy of the results \cite{pharr1998measurement}. The indentation depths were kept below 20\% of the films' thickness. For each fluorine concentration, at least 8 indentations were measured and averaged. For all indentations, the same trapezoidal loading function was used, with load/unload rate of 2 $\mathrm{mN/min}$ and maximum load of 1 $\mathrm{mN}$ (following the methodology of Yao et al. \cite{yao2004structural}, and a 2-second plateau hold time at the peak load.

For nanoscratch tests, all nanoscratch tests were performed with a loading rate of 10 $\mathrm{mN/min}$ \cite{yao2004structural} and a maximum load of 15 $\mathrm{mN}$ over a lateral displacement of 20 $\mathrm{\mu m}$. Before each scratch, the tip is laterally moved by 10 $\mathrm{\mu m}$ to one side for tilt correction, and the scratch moves to -10 $\mathrm{\mu m}$. For each F-DLC film condition, at least 5 scratch tests were performed with the same load function, and during the analysis, all friction curves were reviewed to ensure consistency, and one representative curve was selected and plotted for clarity.

\section{Supplementary Materials}
\subsection{Spectroscopic Ellipsometry of DLC and F-DLC Films}

To ensure that the DLC and F-DLC film thickness does not influence the optical contact angle results, it is important to confirm that the thickness exceeds the critical value of 200 nm, as shown in Figure 1c of the main text. The Filmetrics F20 reflectometer provided a fast and convenient method for determining film thickness, validated using SEM imaging and variable-angle spectroscopic ellipsometry (VASE, J.A. Woollam). Our results confirm that the DLC reference film thickness is 260 nm.

The depolarization values from ellipsometry measurements are typically reported as wavelength-dependent values $\Psi$ and $\Delta$, related to the complex-valued reflectance coefficients \cite{Schweizer2005, Fujiwara2007}:
\begin{equation}
    \frac{r_p}{r_s} = \tan(\Psi) e^{i\Delta}
\end{equation}

where $r_p$ is the complex-valued reflectance of the sample for $p$-polarized light, and $r_s$ is the complex-valued reflectance for $s$-polarized light. And by fitting the $\Psi$ and $\Delta$, we could obtain the optical properties (refractive index) and the thickness of the material

\begin{figure}[ht]
      \centering
        \includegraphics[width=\textwidth]{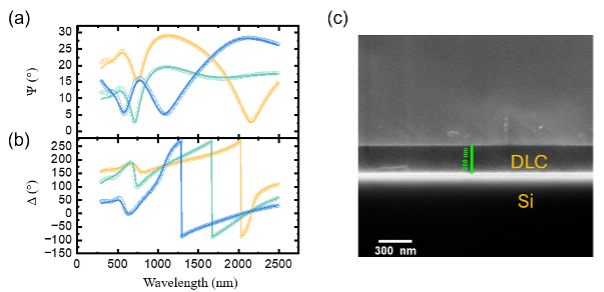}
     \hfill
    \caption{(a) (b) Experimental $\Psi$ and $\Delta$ data (symbols) and corresponding fits (lines) at incidence angles of 50°, 60°, and 70° for reference DLC film grown on top of silicon substrate. (c) SEM image of the cross section of the reference DLC film.}
    \label{fig:FigS1}
\end{figure}

We plotted experimental $\Psi$ and $\Delta$ versus wavelength for DLC films grown on top of silicon substrate in Figure \ref{fig:FigS1} (a) and (b). The fitted result matches the experimental data well. The fitted thickness of the top DLC layer is 260 nm, which is in good agreement with the reflectometer data and the SEM image of the cross section (Figure \ref{fig:FigS1}(c)). And it is thicker than the critical value of 200 nm so the pure DLC film should not affect the optical contact angle measurements.

We also did ellipsometry measurements on the 22.19\% F-DLC sample. However, the ellipsometry fitting for the F-DLC samples was more challenging due to the presence of a DLC interlayer between the top F-DLC layer and the bottom silicon substrate. Typically, when fitting ellipsometric data for a stacked structure like this, it is essential to have a good estimated thickness of each layer in advance. Without an accurate thickness estimation, there are four variables—the thickness and the refractive index of the two layers—changing simultaneously, making it difficult to achieve a unique and accurate fit.

The refractive index derived from ellipsometry is useful for understanding the microstructure and bonding properties of the DLC films, including the $sp^2/sp^3$ ratio. And it plays a significant role in optimizing the films' performance for applications such as optical coatings and biomedical implants.

We used Cody-Lorentz oscillator, developed by Ferlauto, et al \cite{Ferlauto2002}, that is designed for amorphous materials to fit the DLC ellipsometry data. The Cody-Lorentz oscillator is defined as \cite{Woollam2002}:

\begin{equation}
\varepsilon_{Cody-Lorentz}(E) = \varepsilon_1(E) + i\varepsilon_2(E),
\end{equation}

\begin{equation}
\varepsilon_2(E) =
\begin{cases}
\frac{E_1}{E} \exp \left( \frac{E - E_t}{E_u} \right), & 0 \leq E \leq E_t \\[12pt]
\frac{(E - E_g)^2}{(E - E_g)^2 + E_p^2} * \frac{A E_0 \Gamma E}{(E^2 - E_0^2)^2 + \Gamma^2 E^2}, & E > E_t
\end{cases}
\end{equation}

\begin{equation}
\varepsilon_1(E) = \frac{2}{\pi} P \int_{0}^{\infty} \xi \frac{\varepsilon_2(\xi)}{\xi^2 - E^2} d\xi,
\end{equation}
Where $\varepsilon_1$ is the real part of the dielectric function, $\varepsilon_2$ is the imaginary part of the dielectric function, $A$ is the unitless Lorentz oscillator amplitude, $E_0 (eV)$ the peak transition energy, $E_g (eV)$ the optical band gap energy, and $\Gamma (eV)$ is oscillator width. $E_t (eV)$ is the transition energy between the Urbach tail and band-to-band transitions, $E_p (eV)$ is the transition energy that separates the beginning manners of absorption from the Lorentzian behavior and $E_u (eV)$ is the weak Urbach absorption energy \cite{AbdelWahab2022}. The above equations represent a Hilbert transform which ensures the real and imaginary parts of the dielectric function are Kramers-Kronig consistent \cite{Carcione2019}.

We can obtain the complex refractive index by summing the contributions of the Cody-Lorentz oscillator and $\epsilon_\infty$, which is the value of the dielectric function at frequencies much higher than the highest-frequency oscillator:
\begin{equation}
\tilde{n}^2 = (n + i\kappa)^2 = \varepsilon_{\infty} + \sum_{m} \varepsilon_{Cody-Lorentz}
\end{equation}

\begin{figure}[ht]
      \centering
        \includegraphics[width=0.8\textwidth]{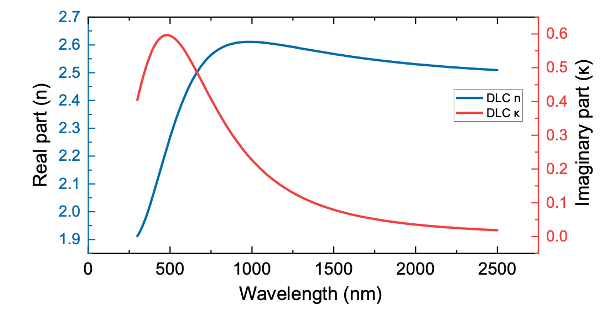}
     \hfill
    \caption{Complex refractive index of our DLC reference sample.}
    \label{fig:FigS2}
\end{figure}

Using the Cody-Lorentz oscillator model, we fitted the raw ellipsometry data to determine the wavelength-dependent refractive index and extinction coefficient of the DLC reference sample (Figure \ref{fig:FigS2}). The fit parameters are listed in Table \ref{tab:DLC_parameters}.

\begin{table}[h]
    \centering
    \caption{Cody-Lorentz Oscillator Parameters for DLC Sample}
    \centering
    \renewcommand{\arraystretch}{1.2}
    \begin{tabular}{|c|c|c|c|c|c|c|c|c|}
        \hline
        & $\varepsilon_{\infty} (eV)$ & $A$ & $E_n (eV)$ & $\Gamma (eV)$ & $E_g (eV)$ & $E_p (eV)$ & $E_t (eV)$ & $E_u (eV)$ \\
        \hline
        \textbf{DLC} & 2.2568 & 44.846 & 2.7323 & 3.8714 & 0 & 6.3109 & 0 & 0 \\
        \hline
    \end{tabular}
    \label{tab:DLC_parameters}
\end{table}

\subsection{$sp^2/sp3$ ratio in the XPS C1s deconvolution}
\begin{figure}
      \centering
        \includegraphics[width=0.8\textwidth]{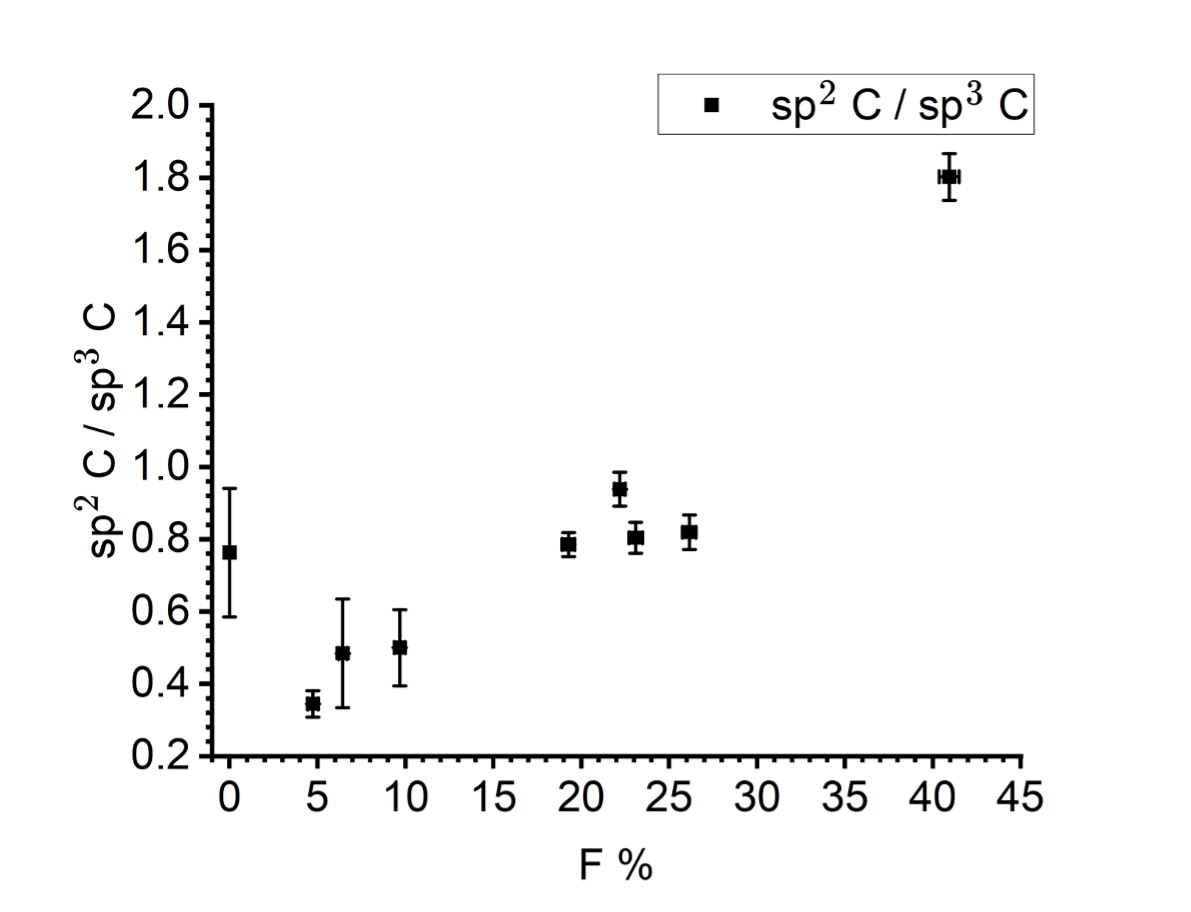}
     \hfill
    \caption{XPS C1s deconvolution shows that sp2/sp3 ratio increases as F\% increases.}
    \label{fig:FigS3}
\end{figure}

Figure \ref{fig:FigS3} illustrates how the XPS C1s deconvolution indicates an increasing $sp^2/sp^3$ ratio with rising fluorine content.

\subsection{SEM Imaging of F-DLC Surfaces}
\begin{figure}[hbt!]
      \centering
        \includegraphics[width=\textwidth]{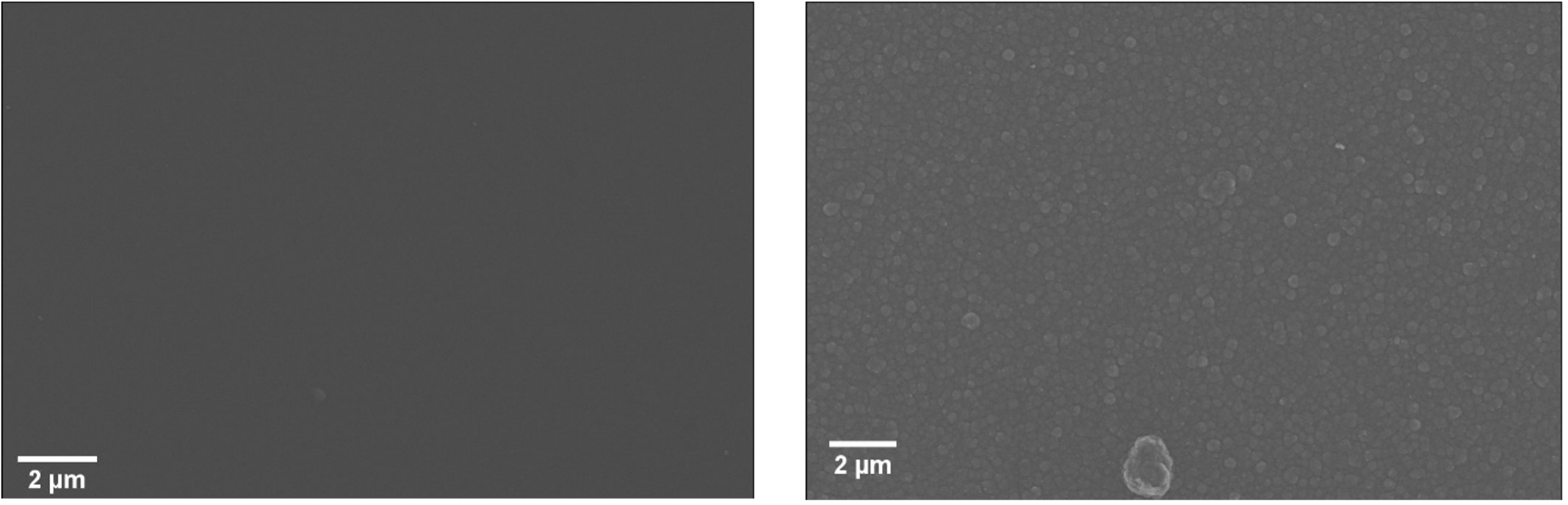}
     \hfill
    \caption{Top-view SEM of (a) F\%=6.43\% and b) F\%=26.14\% F-DLC films.}
    \label{fig:FigS4}
\end{figure}

Top-view SEM images reveal a substantial increase in surface roughness with rising fluorine content in the F-DLC films, confirming the AFM findings in the main text (Figure \ref{fig:FigS4}).

\subsection{XPS C1s Deconvolution Data Table}
Table \ref{tab:xps} presents the deconvolution results of the XPS C1s scan, performed using CasaXPS software.

\begin{table}
    \centering
    \caption{All chemical groups area percentages in C1s deconvolution, from different F\% samples.}
    \begin{tabular}{|c|c|c|c|c|c|c|c|c|}
        \hline
        F\% atomic & F:C gas flow & C-C & C=C & C-CF & C-F & C-F2 & C-O & C=O \\
        percentage & ratio & (\%) & (\%) & (\%) & (\%) & (\%) & (\%) & (\%) \\
        \hline
        0 & 0 & 53.16 & 40.55 & 0 & 0 & 0 & 4.79 & 1.5 \\
        4.73 & 0.4 & 66.03 & 22.76 & 2.93 & 1.01 & 0 & 4.04 & 3.22 \\
        6.43 & 0.5 & 61.22 & 29.66 & 3.78 & 0.87 & 0 & 1.91 & 2.57 \\
        9.67 & 0.667 & 58.75 & 29.37 & 2.02 & 1.33 & 0 & 4.89 & 3.64 \\
        22.19 & 1 & 35.45 & 33.26 & 8.12 & 5.92 & 1.78 & 4.01 & 11.47 \\
        19.27 & 1.5 & 44.34 & 34.82 & 4.76 & 3.31 & 0.82 & 2.36 & 9.58 \\
        23.1 & 1.7 & 40.75 & 32.77 & 5.91 & 4.58 & 1.53 & 3.23 & 11.24 \\
        26.14 & 1.85 & 38.34 & 31.42 & 6.58 & 5.56 & 1.68 & 2.97 & 13.44 \\
        40.93 & 2 & 20.52 & 36.98 & 9.24 & 10.53 & 3.56 & 10.26 & 8.9 \\
        \hline
    \end{tabular}
    \label{tab:xps}
\end{table}

\section{Acknowledgements}
This work was supported by the Advanced Materials Industrial Consortium (AMIC) of the Wisconsin Materials Research Science and Engineering Center (MRSEC, NSF DMR-1720415). Work by AT was supported by the U.S. Department of Energy, Office of Science, Basic Energy Sciences under Award DE-SC0023873. Work by RV and JTC are supported by the U.S. Department of Energy, Office of Science, Basic Energy Sciences under Award DE-SC0020313. The authors gratefully acknowledge use of facilities and instrumentation in the UW-Madison Wisconsin Center for Nanoscale Technology. The Center (wcnt.wisc.edu) is partially supported by the Wisconsin Materials Research Science and Engineering Center (NSF DMR-2309000) and the University of Wisconsin-Madison. The authors are grateful to Mikhail Kats for helpful discussions.

\appendix

\clearpage
\bibliographystyle{elsarticle-num} 
\bibliography{F-DLC}

\nocite{schweizer2005handbook,fujiwara2007spectroscopic,ferlauto2002analytical,woollam2002guide,abdel2022spectroscopic,carcione2019kramers}




\end{document}